Maine Business School &
School of Computing and
Information Science

5723 DP Corbett Building
Orono, Maine 04469
Tel:  207-581-1968
Fax:  207-581-1930
umaine.edu/business

# AI Ethics and Ordoliberalism 2.0: Towards A 'Digital Bill of Rights'

Manuel Wörsdörfer[1]

**Abstract:** This article analyzes AI ethics from a distinct business ethics perspective, i.e., 'ordoliberalism 2.0.' It argues that the ongoing discourse on (generative) AI relies too much on corporate self-regulation and voluntary codes of conduct and thus lacks adequate governance mechanisms. To address these issues, the paper suggests not only introducing hard-law legislation with a more effective oversight structure but also merging already existing AI guidelines with an ordoliberal-inspired regulatory and competition policy. However, this link between AI ethics, regulation, and antitrust is not yet adequately discussed in the academic literature and beyond. The paper thus closes a significant gap in the academic literature and adds to the predominantly legal-political and philosophical discourse on AI governance. The paper's research questions and goals are twofold: First, it identifies ordoliberal-inspired AI ethics principles that could serve as the foundation for a 'digital bill of rights.' Second, it shows how those principles could be implemented at the macro level with the help of ordoliberal competition and regulatory policy.

## 1. Introductory Remarks

Dozens of AI ethics initiatives and governance documents have emerged over the past few years, starting with the U.S. National Science and Technology Council's 'Preparing for the Future of AI' and the E.U. Digital Charter in 2016. The latest examples include the Biden-Harris Administration's 'Blueprint for an AI Bill of Rights' and the White House's 'Ensuring Safe, Secure, and Trustworthy AI Principles' (2023).[2]

Attard-Frost et al. (2022) reviewed 47 AI guidelines and identified the FAST principles: fairness, accountability, sustainability, and transparency. According to the authors, fairness and accountability themes can be found in 20 out of 47 guidelines, sustainability in 11, and transparency in 14. A similar study was conducted by Fjeld et al. (2020), who analyzed 36 prominent AI principles documents published by different actor groups – i.e., governments, intergovernmental organizations, companies, professional associations, advocacy groups, and multi-stakeholder initiatives – and identified eight common themes: privacy (present in 97% of all documents in the database), accountability (97%), safety and security (81%), transparency and explainability (94%), fairness and non-discrimination (100%), human control of technology (69%), professional responsibility (78%), and promotion of human values (69%). According to Leslie et al. (2021), the fundamental AI ethics principles can be summarized as human dignity, freedom and autonomy, harm prevention, non-discrimination, equality and diversity, transparency and explainability of AI systems, data protection and the right to privacy, accountability and responsibility, and democracy and the rule of law.

---

[1] *Address for correspondence:* Dr. Manuel Wörsdörfer, Assistant Professor of Management and Computing Ethics, Maine Business School & School of Computing and Information Science; University of Maine; E-Mail: manuel.woersdoerfer@maine.edu.

[2] Other notable examples include the Future of Life Institute's 'Asilomar AI Principles' (2017), UNI Global Union's 'Top 10 Principles for Ethical AI' (2017), Council of Europe's 'European Ethical Charter on the Use of AI in Judicial Systems' (2018), European Commission's 'AI for Europe' (2018), Germany's 'AI Strategy' (2018), 'Beijing AI Principles' (2019), G20's 'AI Principles' (2019), High-Level Expert Group on AI's 'Ethics Guidelines for Trustworthy AI' and 'Policy and Investment Recommendations' (2019), IEEE Global Initiative's 'Ethically Aligned Design' (2019), OECD's 'Principles on AI' (2019), Global Partnership on AI (2020), E.U.'s 'White Paper on AI' (2020), U.K.'s National AI Strategy (2021), E.U.'s 'Proposal for a Regulation on a European Approach to Artificial Intelligence' (2021), and Khanna's 'Internet Bill of Rights' (2022). Furthermore, several (inter-)national standardization efforts are underway, e.g., by standard-developing organizations such as ISO, IEC, NIST, CEN, and CENELEC (AlgorithmWatch, 2023; Hagendorff, 2020; Jobin & Vayena, 2019; Smuha, 2019).

While AI ethics (initiatives) play an essential role in motivating morally acceptable professional behavior and prescribing fundamental duties and responsibilities of computer engineers and can, therefore, bring about fair(er), safe(er), and (more) trustworthy AI applications (Laux et al., 2023a; Leslie, 2019; Rieder et al., 2021; Rubenstein, 2021), they also come with various shortcomings: One of the main concerns is that the proposed AI guiding principles are often too abstract, vague, flexible, or confusing (i.e., overlapping or contradicting) (Mittelstadt, 2019) and lack proper implementation guidance (Morley et al., 2021). Consequently, there is often a gap between theory – i.e., principles – and practice – i.e., implementation – resulting in a lack of practical operationalization by AI industry players. Other critics point out the potential trade-off between ethical principles and corporate interests (i.e., profit and shareholder value maximization) and the possible use of those initiatives for ethics-washing or window-dressing purposes (Metzinger, 2019). Furthermore, most AI ethics guidelines are soft-law documents that lack adequate governance mechanisms and do not have the force of binding law, further exacerbating white or greenwashing concerns (see, for the latest examples, the Biden-Harris Administration's 'Blueprint for an AI Bill of Rights' [White House, 2023a; Hine & Floridi, 2023] and 'Ensuring Safe, Secure, and Trustworthy AI Principles' [White House, 2023b] and the formation of the industry body 'Frontier Model Forum,' supported by Anthropic, Google, Microsoft, and OpenAI [OpenAI, 2023a]). Lastly, there is also possible regulatory or policy arbitrage, so-called jurisdiction or 'ethics shopping' to countries with laxer standards and fewer constraints, e.g., offshoring to countries with less stringent mandatory requirements for AI systems (Attard-Frost et al., 2022; Daly et al., 2021; Morley et al., 2021; Rubenstein, 2021).

This article analyzes these AI ethics and governance issues (Dubber & Pasquale, 2020) from a distinct business ethics, i.e., a revised ordoliberal perspective, so-called 'ordoliberalism 2.0' (Wörsdörfer, 2020, 2022b), a perspective currently lacking in the academic literature. It argues that the ongoing political discourse on (generative) AI relies too much on corporate self-regulation and voluntary codes of conduct. Thus, it lacks adequate governance mechanisms, including monitoring, enforcement, and sanctioning. To address the above issues, the paper suggests not only introducing hard-law regulation with a more effective oversight structure (this would imply, among others, a fundamental overhaul of the proposed E.U.'s Artificial Intelligence Act [AIA] [see Section 4 below]) but also merging already existing AI ethics guidelines with an ordoliberal-inspired regulatory framework and competition policy. However, this link between AI ethics, regulation, and antitrust is not yet adequately discussed in the academic literature and beyond (this is especially the case for competition policy, which could help address the increasing power asymmetries in the digital economy, e.g., between big tech, consumers, and small and medium-sized companies). The paper thus closes a significant gap in the academic literature and adds to the predominantly legal-political and philosophical discourse on AI ethics and governance.

This paper's research questions and goals are twofold: First, it identifies and analyzes ordoliberal-inspired AI ethics principles that could serve as the foundation for a 'digital bill of rights.'[3] Second, it shows how those principles could be implemented at the macro (i.e., nation-state) level with the help of ordoliberal competition and regulatory policy.

Building on our previous work on ordoliberalism (Wörsdörfer 2013b, 2022c), big tech and antitrust, with a particular focus on the E.U.'s Digital Markets Act (DMA) (Wörsdörfer, 2020, 2021, 2022a, 2022b, 2023a), and the AIA (Wörsdörfer, 2023b, 2023c), the paper shows that AI ethics could benefit significantly from revised ordoliberalism as this business-ethical concept provides valuable lessons for regulating the digital economy in the 21$^{st}$ century (Wörsdörfer, 2022b).

---

[3] I.e., respect for human rights, data protection and the right to privacy, harm prevention and beneficence, non-discrimination and freedom of privileges, fairness and justice, transparency and explainability of AI systems, accountability and responsibility, democracy and the rule of law, and environmental and social sustainability.

Especially the link between AI ethics frameworks (i.e., regulatory policy), on the one hand, and big tech and antitrust (i.e., competition policy), on the other, makes ordoliberalism 2.0 vital for the current discussion on AI technologies' ethical issues and implications.

The paper is structured as follows: Section 2 provides an overview of the key characteristics of ordoliberalism. Section 3 identifies fundamental ordoliberal AI ethics principles and the role they could play in the digital economy. Section 4 analyzes possible ways to realize the principles identified in the previous section(s), focusing mainly on regulatory (i.e., AI legislation) and competition policy (i.e., big tech and antitrust). The paper ends with a summary of its main findings and an outlook on future research questions.

## 2. Ordoliberalism

Ordoliberalism is a business-ethical concept that has paved the way for Germany's – and the E.U.'s – social market economy implemented after World War II. It attempts to bridge the gap between moral (i.e., social justice, human rights) and economic (i.e., competition, market freedom) imperatives. Its primary goal is to establish an economically efficient and, at the same time, humane socio-economic order – one that can protect the Kantian values of freedom, autonomy, and dignity (Oppenheimer, 1933; Röpke, 1944/1949; Rüstow, 2001; Wörsdörfer, 2013b). Two schools of economic thought must be distinguished within classical ordoliberalism – the Freiburg School of Law and Economics and 'sociological neoliberalism.' The founders of the Freiburg School, an interdisciplinary research group at the University of Freiburg, were the economist Eucken and the two legal scholars Böhm and Großmann-Doerth; the leading representatives of sociological neoliberalism were Rüstow and Röpke (Wörsdörfer, 2022c). Besides classical ordoliberalism, there is also contemporary ordoliberalism (Feld & Köhler, 2011; Goldschmidt, 2002, 2007; Goldschmidt & Wohlgemuth, 2008; Häußermann & Lütge, 2022; Vanberg, 2004, 2005, 2008b, 2013; Wörsdörfer, 2020), which bears considerable resemblances to constitutional economics à la Buchanan and evolutionary liberalism à la Hayek.

Amongst the most famous ordoliberal catchwords is the differentiation between *Ordnungspolitik* and *Prozesspolitik* (Eucken, 1950/1965, 1952/2004, 1999, 2001): According to the 'Freiburg imperative,' regulatory or ordering policy is favored, which means that the government as a legislator and rule-maker – and not as a significant economic player – is responsible for setting, preserving, and enforcing the regulatory framework. The government should restrict itself to economic policies that frame or define the general terms and conditions under which market transactions occur. In other words, the government must focus solely on establishing, monitoring, and enforcing the 'rules of the game' instead of steering, influencing, or intervening in market processes and the play itself. The overall goal of regulatory policy is to implement a competitive socio-economic order capable of safeguarding freedom, autonomy, citizen sovereignty, and dignity (this presupposes a constitutional state based on the rule of law).

Eucken's Principles of Economic Policy – and his Constituent and Regulating Principles (see Table 1 below and Eucken, 1952/2004) – demand not only the disempowerment of socio-economic lobbying and pressure groups; they also require a 'market-conforming' regulatory policy (Röpke, 1942, 1944/1949) – one that does not interfere in the market and price mechanism – and the dismissal of a market-non-conforming process policy (Eucken, 1952/2004).

| Constituent Principles | Regulating Principles | Principles of Economic Policy |
|---|---|---|
| Competitive Economy | Correction of Market Power | Limiting Power of Rent-Seekers |
| Monetary Policy & Price Stability | Income Redistribution (Distributive Justice) | Focus on Regulatory Policy |
| Open Markets | Correction of Negative Externalities | |
| Property Rights | Correction of Abnormal Supply Reactions | |
| Freedom of Contracts | | |
| Principle of Liability | | |
| Regularity of Economic Policy | | |
| Interdependency of Principles | | |

*Table 1: Eucken's Ordoliberal Principles[4]*

The latter is rejected for several reasons: Process policy is considered by Eucken and other ordoliberals as a form of 'privilege-granting policy.' It is mainly based on ad-hoc and case-by-case decisions and enables arbitrary and selective interventions in the economic 'game of catallaxy' (Hayek, 1973). It thus lacks two crucial features of an ordoliberal economic policy – predictability and long-term orientation. Most importantly, however, it opens the doors for special interest groups to exert influence on the legislative decision-making process: That is, process policy is more likely to be prone to the power of rent-seeking or lobbying groups – due to a more significant regulatory load, and the existence of a higher discretionary leeway for decision-making. It thus goes hand in hand with a considerable lack of transparency as many debates and decisions take place behind closed doors – and a lack of accountability and democratic legitimacy – since interest groups represent only a fraction of society and are seldom directly and democratically elected (besides, process policy also tends to weaken or undermine constitutional checks and balances). In sum, this form of particularistic policy jeopardizes the nation's wealth – due to granting costly and exclusive privileges to special interest groups – and undermines personal freedom – due to the increased politico-economic powers of rent-seekers.

---

[4] Eucken, 1952/2004; Wörsdörfer, 2022b.

According to (classical) ordoliberalism, it is essential to discuss the relationship between markets and states and clearly define the government's tasks and the limits of its responsibilities. The ideal ordoliberal state is a strong and independent constitutional state (Röpke, 1942, 1944/1949, 1950; Rüstow, 1955), a state that stands above special interest groups and serves as a 'market police' (Röpke, 1942; Rüstow, 1957, 2001), 'ordering power,' and 'guardian of the competitive order' (Eucken, 1952/2004; Röpke, 1944/1949). The state should ideally be able to fend off special interest groups, keep to the principles of neutrality and impartiality, and confine itself to regulatory policy. The underlying liberal ideals are equality before the law (i.e., the rule of law), freedom of privileges, and the principle of non-discrimination (Böhm, 1966/1980; Vanberg, 2008a) – similar to that of Buchanan and Hayek's constitutional economics (Brennan & Buchanan, 1985/2000; Buchanan, 1975/2000; Buchanan & Congleton, 1998; Buchanan & Tullock, 1962/1999; Congleton, 2013; Vanberg, 2008b; Wörsdörfer, 2020).

Overall, the ordoliberals are searching for an integrative third way (Oppenheimer, 1933; Rüstow, 2001) between the social Darwinism of laissez-faire capitalism (i.e., Scylla of paleo-liberalism) and totalitarian collectivism and Hobbes' Leviathan (i.e., Charybdis of socialism). They refer to this type of socio-economic model as ordoliberalism (with 'ordo' simply meaning socio-economic and political order [Eucken, 1950/1965]), social liberalism, or economic humanism (Röpke, 1944/1949; Rüstow, 2001).

Figure 1 summarizes the main principles of 'ordoliberalism 2.0' derived from the analysis in this section (Wörsdörfer, 2022b).

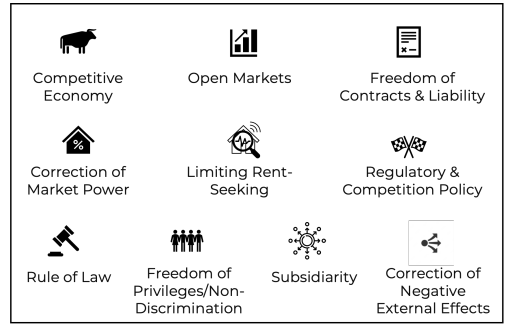

*Figure 1: Principles of Ordoliberalism (2.0)*

### 3. Ordoliberal AI Ethics Principles

Based on Section 2 and our previous work (Wörsdörfer, 2020, 2021, 2022a, 2022b, 2022c, 2023a, 2023b, 2023c), we can identify several ordoliberal principles that are of particular relevance to AI ethics:

- Respect for human rights,
- Data protection and the right to privacy,
- Harm prevention and beneficence,
- Non-discrimination and freedom of privileges,
- Fairness and justice,
- Transparency and explainability of AI systems,
- Accountability and responsibility,
- Democracy and the rule of law, and
- Environmental and social sustainability.

The subsequent sections provide an overview of each of those values, i.e., a brief explanation of the norm's role within ordoliberalism, followed by a detailed assessment of the importance of the ordoliberal-inspired principle in an AI context.

*Respect for Human Rights*

Eucken's 'program of liberty' stands in the tradition of (neo-)Kantian ethics, inspired by the work of his father Rudolf, a philosophy professor and Nobel Prize laureate, and Husserl, founder of phenomenology and close friend of the Eucken family (Klump & Wörsdörfer, 2011). Like Kant,

Eucken considers humans as free, rational, and autonomous agents who are capable of giving themselves their own moral laws and setting and pursuing their own ends (GMM 4:421; CPrR 5:20/5:21). Because humans can guide their conduct by reason and morality they possess an intrinsic or inner moral value, i.e., dignity (GMM 4:435), which makes them valuable above all else (i.e., things that can be exchanged have a price; ends in themselves, however, have dignity [GMM 4:434]). As such, humans as *Moralwesen* and *Vernunftwesen* possess an innate, inalienable, and inviolable self-worth that must always be respected. In particular, Kant's 'Humanity or End-in-itself-Formula' (GMM 4:429) commands that "you use humanity, whether in your own persona or in the person of any other, always at the same time as an end, never merely as a means." It implies respecting other people's moral and rational (human) agency and rules out treating others as mere instruments, objects, or tools.

According to Eucken and other ordoliberals, freedom is a constituent of human existence. It is founded on the Kantian concepts of self-determination, self-legislation, and human dignity (Eucken, 1948a; see Böhm [1966/1980] for more information on his Kantian idea of 'private law society'). The overall goal is to overcome a person's immaturity and reach a state of personal enlightenment and emancipation. It is noteworthy that Eucken's understanding of liberty is not anarchistic; a person's freedom finds its limits where another person's sphere of liberty begins. Moreover, there is no absolutization of freedom in Eucken's writings; on the contrary, several societal constraints and moral limitations exist, e.g., in the form of Judeo-Christian ethics, which emphasizes the importance of solidarity and justice, to name a few. For Eucken and other ordoliberals, liberty thus goes hand in hand with an encompassing sense of responsibility – towards oneself and others (i.e., a double form of accountability: individual *and* social responsibility [Eucken, 1953]). Finally, Eucken and others strive to connect negative and positive and economic and political freedom: Liberty is consequently not just limited to negative freedom (i.e., freedom from coercion and [government] interventions) – it also incorporates a positive notion of freedom strongly related to the Kantian idea of autonomy (i.e., freedom to act and freedom to achieve) (Berlin, 1969). In addition, freedom is not just restricted to the economic sphere, that is, free markets and economic freedom; it is also essential in the political arena (Eucken, 1946/1999a): i.e., inalienable fundamental human rights, guaranteed by a constitutional state and the rule of law, enable the free development of the individual (Eucken, 1952/2004). In short, political freedom is interconnected with human rights and dignity; economic freedom, on the other hand, relates to consumer sovereignty and 'coordinating' individual plans via competitive markets instead of authoritarian or collectivist 'subordination' (Eucken, 1952/2004). Lastly, it is worth noting that Eucken's understanding of freedom is incompatible with various socio-economic and political dependencies, oppression, exploitation, and all totalitarian ideologies and authoritarian regimes (in fact, he was part of the resistance movement [in the wider sense] during the Nazi era) (Eucken, 1932a, 1932b, 1948a; Wörsdörfer, 2022c). According to him, freedom is threatened in three ways (Eucken, 1952/2004): first, it is threatened by the private power of large corporations and especially cartels and trusts; second, it is threatened by 'the collective' (i.e., the risk of 'group anarchy' and collective ideologies), and third, it is threatened by the power of the 'big government.' The best way to address these threats is with the help of an ordoliberal-inspired, highly competitive regulatory policy based on Eucken's Principles of Economic Policy and his Constituent and Regulating Principles (see above).

In the context of AI, Eucken's Kantian program of liberty and human rights requires a human-centered approach to AI and preserving human agency, control/oversight, and responsibility in the digital economy – as one major precondition for guaranteeing fundamental rights and human dignity (i.e., the freedom and autonomy of individuals). Human control of AI technologies requires, among others, a human review of automated decisions and the ability to opt out of computerized decisions (Fjeld et al., 2020). It also implies evaluating the societal impacts of AI systems and their effects on human agency (Kazim & Koshiyama, 2021) as well as the promotion of human values, including human well-being and flourishing, access to technology, and leveraging technology for the benefit of society (Fjeld et al., 2020).

According to Prabhakaran et al. (2022) and others, such a Kantian-inspired, ordoliberal human rights framework can serve as the foundation for building 'value-aligned AI systems.' Human rights – understood here as moral claims and entitlements (Wettstein, 2009a, 2009b, 2010a, 2010b, 2012a, 2012b) – play a significant role in AI (ethics) given that such technologies directly impact people's right to …

- *Life, liberty, and well-being* (e.g., AI systems denying people employment and housing opportunities and loans),

- *Security, data protection, and privacy* (e.g., indiscriminate and suspicionless mass surveillance by 'big government' and 'big business' undermining privacy, encryption, and anonymization rights),
- *Freedom from discrimination* (e.g., AI technologies disproportionately negatively impacting people of color; such ethnic discrimination includes, among others, predictive policing and racial profiling by law enforcement agencies]),
- *Freedom of speech and expression* (e.g., chatbots and deepfakes spreading mis-/disinformation [New York Times, 2023a, 2023b] and other harmful content [OpenAI, 2023b; Responsible AI Collaborative, 2023], including dangerous/hate speech, thereby affecting public discourses in democratic and rule-of-law societies]),
- *Healthcare* (e.g., biases in algorithmic systems [i.e., 'algorithmic unfairness'] denying people health insurance or medical treatments), and
- *Education and science* (e.g., AI technologies denying college admissions and financing; current AI R&D is also dominated by Western researchers and practitioners and Western value systems and institutions]; it thus lacks diversity, equity, and inclusion [Ajayi et al., 2023; Prabhakaran et al., 2022]).[5]

A key advantage of the Kantian-ordoliberal human rights framework is that it is already part of a legal regime and set of government instruments (see the Universal Declaration of Human Rights and the International Covenants on Civil and Political Rights and Economic, Social, and Cultural Rights). Additionally, it is increasingly reflected in business and cultural practices (see the U.N. Guiding Principles on Business and Human Rights emphasizing human rights due diligence) and an essential element of a global social movement focusing on empowering people and advocating for socio-economic rights and cultural change. As such, the human rights framework possesses intercultural and cross-cultural validity (i.e., human rights are no longer culturally parochial or time-bound and find cross-cultural support) and can help overcome existing biases in AI research and practice. That is, the human rights framework can potentially incorporate marginalized and vulnerable stakeholder perspectives, including but not limited to indigenous peoples, BIPOC (black, indigenous, and people of color), and people from developing countries. This, however, requires fostering stakeholder engagement and dialogue and inclusion and participation in AI R&D and the business world (Ajayi et al., 2023; Prabhakaran et al., 2022; see also Section 4 below).

---

[5] O'Neil (2016) classifies AI systems as 'weapons of math destruction' since they negatively impact the marginalized and vulnerable parts of society, i.e., low-income people and ethnic minorities. That is, they often lead to more discrimination, racism, and prejudices, e.g., due to biased software and data, thereby increasing inequality, deepening the social divide, and negatively impacting democracy and the rule of law. AI systems also reinforce negative feedback loops and vicious circles, e.g., in the form of poverty traps. Lastly, victims of AI discrimination and racism have (almost) no means to file a complaint and mitigate their harms and adverse impacts (see also Deutscher Ethikrat, 2023; Institute for Human Rights and Business, 2016).

A potential disadvantage of a Kantian human rights framework is that it might amplify the existing anthropocentric worldview, potentially neglecting the interests of non-human life. This criticism could be addressed with Eucken's focus on environmental sustainability and correcting adverse external effects (see the section on socio-environmental sustainability and Section 4).

*Data Protection and the Right to Privacy*

Closely related to the first principle, ordoliberals would also insist on a right to privacy and data protection. Both are seen here as a human right and a specific form of autonomy and individual freedom (see for the following, Wörsdörfer, 2018).

Privacy as a human right – see Article 12 of the Universal Declaration of Human Rights or Article 17 of the International Covenant on Civil and Political Rights – is an essential part of what it means to be human. It is a core condition of being a free person and, therefore, crucial for human freedom and dignity. It also enables liberation from societal constraints. That is, denying privacy severely restricts one's freedom of choice, and deprivation or invasion of privacy crushes any temptation to deviate from rules or norms. Privacy is also essential for human happiness and well-being since it directly relates to many attributes typically associated with the quality of life, such as creativity, exploration, and intimacy. It allows people to be themselves and wear their 'private faces' (i.e., privacy as the 'right to be let alone' [Warren & Brandeis, 1890; for more information on the relationship between Eucken and Brandeis, see Wörsdörfer, forthcoming]). If people did not have a minimum level of privacy, they would be forced to wear their 'public faces' at all times and under all circumstances, which would have long-lasting and damaging effects on their psychological well-being. Privacy is thus valuable because it allows us to develop and maintain loving, trusting, caring, and intimate relationships with family and friends, fostering social capital within society. Furthermore, privacy is necessary for an individual to blossom as a person and for individual growth: It promotes and stimulates intellectual activities by allowing people to be creative and to grow spiritually. It also allows people to explore new frontiers and experiment with the boundaries of rules and norms – a significant precondition for economic and technological progress in society. Lastly, the ethical importance of privacy rests in being a critical pre-condition for free speech and freedom of expression (i.e., privacy as informational self-determination and sovereignty).

The key privacy dimensions – from an ordoliberal point of view – can be summarized as follows (Wörsdörfer, 2018):

- Integrity and dignity (i.e., privacy as a guarantor of human dignity),
- Personhood and identity (i.e., privacy as [citizen] sovereignty, autonomy, and self-determination),
- Intimacy and anonymity (i.e., privacy as 'right to be let alone'),
- Control over data and information (i.e., privacy and information control),
- Limited access to self (i.e., privacy as 'zone of inaccessibility'), and
- Freedom of speech and expression (i.e., privacy and communication freedoms).

In the context of AI, privacy as a human right (i.e., autonomy and freedom) implies a significant limitation of arbitrary mass surveillance and spying – both from government agencies and tech companies – (i.e., data stewardship and minimization [Kazim & Koshiyama, 2021]), informational self-determination and sovereignty (i.e., giving free, prior, and informed consent and having an opt-in model instead of the current opt-out and notice and choice paradigm), control over data use and the ability to restrict the processing of data, the right to rectification, the right to correct (i.e., having access to personal information and being able to correct or amend database records), the right to erasure, privacy by design/default, data security (i.e., the right to anonymization and encryption), and data protection laws, similar to the General Data Protection Regulation and the California Consumer Privacy Act/California Privacy Rights Act (Fjeld et al., 2020).

*Harm Prevention and Beneficence*

Harm prevention or non-maleficence in the context of AI relates, inter alia, to safety and security (see for the lack of safety in chatbots and other generative AI [models], OpenAI, 2023b; Zou et al., 2023). Key criteria in this regard are the technological robustness of AI systems, the prevention of the malicious use of AI technologies, the reliability and reproducibility of AI research methods and applications, the availability of fallback plans and safe exits, and the consideration of unknown risks (i.e., built-in 'security by design'). Note that societal harm can not only be caused by the misuse, abuse, or poor design but also by negative unintended consequences of AI technologies; special attention must, therefore, be paid to the latter when designing and implementing AI tools (Kazim & Koshiyama, 2021; Leslie, 2019).

AI systems must also be designed to meet the minimum threshold of discriminatory and other forms of non-harm. This requires, among others, that datasets are equitable and that models do not have an inequitable impact and be implemented unbiasedly (Leslie, 2019; for more information, see the sections on non-discrimination and fairness and justice). Similar to cars and airplanes, new safety and security features and standards are also needed. Lastly, new AI legislation must demand mandatory algorithm auditing conducted by independent data auditors and third-party certification and licensing (for more information, see the sections on transparency and accountability and Section 4).

Besides the above negative notion of 'do no harm,' AI developers and operators also have a positive duty to fulfill, i.e., to 'do good.' This notion of beneficence is closely related to the public or common good orientation, as laid out in most engineering and computer ethics textbooks or professional codes of conduct, i.e., promoting the public's health, safety, and well-being (Wörsdörfer, 2018). Ordoliberals, such as Eucken (1953, 1948b, 1950/1996), Röpke (1942, 1944/1949, 1950, 1958/1961), and Rüstow (1945/2001, 1955, 1957), are convinced that (digital) markets and technologies are embedded in a higher, i.e., meta-economic societal order – one that exists 'beyond supply and demand.' The (digital) economy is seen here as a means to an end, whereas Rüstow's 'vital situation' is considered the end in itself. Thus, it must be designed to serve the community, not vice versa (in the context of AI, this implies promoting human flourishing and well-being and leveraging AI systems for the benefit of society [Fjeld et al., 2020]). The (digital) market economy is described as a system that drains or erodes morality and undermines social cohesion; therefore, it requires external, i.e., meta-economic values and bonding forces; (digital) markets (and technologies) themselves cannot generate those necessary moral reserves. Finally, ordoliberals emphasize the importance of market-free sectors and the meta-economic amendment and design of market boundaries, e.g., with the help of ordoliberal 'vital policy' (Wörsdörfer 2013a, 2013b).

*Non-Discrimination and Freedom of Privileges*

With equality before the law and the fight against lobbyism, rent-seeking, and special interest groups, non-discrimination and freedom of privileges play a significant role within ordoliberalism (see Eucken's Principles of Economic Policy and his Constituent and Regulating Principles).

In the context of AI, they relate to avoiding all forms of discrimination, manipulation (e.g., via chatbots and deepfakes), negative (e.g., racial) profiling, and the prevention or minimization of algorithmic biases (i.e., biases in algorithmic decision-making).[6] This requires, among others, representative and high-quality data, as well as fairness, equality, and inclusiveness in both impact and design (Fjeld et al., 2020; Wachter et al., 2021; for more information, see the following section on fairness and justice). Special attention must be paid to vulnerable and marginalized groups, e.g., children, immigrants, and ethnic minorities, and the related problems of possible exclusion and inequality (Ajayi et al., 2023; O'Neil, 2016; Prabhakaran et al., 2022).

Of particular importance from an ordoliberal 2.0 point of view is the preservation of net neutrality (Wörsdörfer, 2018; Wu, 2003): Net neutrality forbids internet service providers – and other gatekeepers (Washington Post, 2023) – to offer discriminatory or tiered online services and charge premium prices for higher-priority routing of internet packets, guaranteeing an open and neutral internet order.

Proponents of net neutrality argue that the internet should be treated as a common good or public utility: The 'Open Internet Order' and its guarantee of equal access to information (services) are not only essential to overcome socio-economic and educational gaps (i.e., the digital and social divides[7]) but are also crucial to protect privacy, freedom of information, and freedom of speech. Thus, net neutrality advocates emphasize the unique role platform neutrality plays within a free, open, and pluralistic society.

Furthermore, ordoliberals point out that abolishing net neutrality and introducing tiered services might hurt and damage small and medium-sized companies since only wealthy and powerful companies could afford the highest level of service (e.g., high-speed internet). As such, tiered services that discriminate between companies and consumers based on their ability and willingness to pay premium prices might discourage competition and innovation and create even more monopolies or oligopolies. The (ordoliberal) proponents of net neutrality thus argue that the Open Internet Order is an essential piece of legislation that guarantees equal (i.e., fair) and open market access to the competitive (digital) economic order – which is primarily based on freedom of discrimination and privileges and the prevention of the monopolization or re-feudalization of the economy.

---

[6] The main reasons for algorithmic biases include the poor selection of training data, especially unrepresentative or incomplete data sets (e.g., relying only on white male U.S. population data or having other cultural or ethnic biases), predictions based on too little data – and thus the impossibility of generalizations –, flawed correlations or not considering the underlying causations, and a lack of diversity among AI developers and data scientists. Note that most computer science teams are dominated by white male Westerners aged 20-40 – so-called 'male AI'; not adequately represented, however, are BIPOC, women, disabled and elderly people, and people from developing countries. Also, note that AI technologies are social artifacts that embed and project AI developers' choices, biases, and values; that is, personal beliefs, opinions, and prejudices, as well as stereotypes and societal biases of computer scientists play a significant role and are reflected in algorithms (i.e., 'bias in, bias out') (Castets-Renard & Besse, 2022; Coeckelberg, 2020; Cowgill et al., 2020; Deutscher Ethikrat, 2023; Kuśmierczyk, 2022; Rubenstein, 2021).

[7] Net neutrality is thus also crucial to realize the ordoliberal concept of 'justice of the starting conditions' (Wörsdörfer, 2013a), discussed in the next section.

Lastly, net neutrality proponents argue that the introduction of tiered internet services would give companies, especially the ones controlling access to specific internet platforms – the so-called gatekeepers of the internet – additional powers to limit and restrict access to non-preferred online content or internet applications, thereby increasing the risk of censorship – which would threaten freedom of information and freedom of speech (Wörsdörfer, 2018).

*Fairness and Justice*

Closely related to the previous principle, freedom of privileges and non-discrimination, ordoliberals also emphasize the importance of fairness and justice considerations, understood here both in a *distributive* – i.e., tackling social injustices and addressing the social question – and *commutative* justice sense – i.e., fairness of rules, procedures, and institutions, as well as 'justice of the starting conditions' and equal opportunities (Wörsdörfer, 2013a).

In the context of AI, four types of AI-related fairness need to be distinguished – data, design, outcome, and implementation fairness (Leslie, 2019). *Data fairness* requires mitigating biases, excluding discriminatory influences, and not generating discriminatory or inequitable impacts on affected individuals and communities. Specifically, AI systems must be trained and tested on adequately representative, relevant, accurate, and generalizable datasets. Thus, data fairness requires representativeness, fit-for-purpose and sufficiency, source integrity and measurement accuracy, timeliness and recency, and relevance, appropriateness, and domain knowledge. *Design fairness* requires that AI systems also have model architectures that do not include target variables, features, processes, or analytical structures (i.e., correlations, interactions, and inferences) that are unreasonable, morally objectionable, or unjustifiable. It thus relates to problem formulation, data pre-processing, feature determination and model building, and evaluating analytical structures. *Outcome fairness* requires that AI systems not have discriminatory or inequitable impacts on the lives of the people they affect. It thus relates to group and individual fairness, includes a fairness position statement, and pays special attention to automation biases. *Implementation fairness* requires that AI systems must be deployed by users sufficiently trained to implement them responsibly and without bias (Leslie, 2019).

From an ordoliberal 2.0 point of view, fairness considerations also relate to open innovation (i.e., open public data repositories, data trusts, or government-operated data sharing arrangements), market fairness (i.e., fair, competitive practices, the promotion of policies that enable small and medium-sized enterprises to compete with large firms more effectively, the reduction of risk of AI platform and data infrastructure monopolies, and the removal of unfair competitive advantages) (Attard-Frost et al., 2022), as well as fostering accessibility (in the sense of 'justice of the starting conditions' and equal opportunities [Wörsdörfer, 2013a]) and promoting inclusion and participation (Kazim & Koshiyama, 2021). Current AI research is often characterized by biases and a lack of professional (i.e., ethnic, gender, etc.) diversity. Specialized education, training, and hiring programs are thus needed to increase diversity and reduce biases in the talent pool; bias reminders are also effective countermeasures, same as representative training data or de-biasing training data (Attard-Frost et al., 2022; Cowgill et al., 2020).

*Transparency and Explainability of AI Systems*

For Eucken and other ordoliberals, transparency and explainability are crucial, especially in the context of economic and political decision-making, e.g., to avoid the various forms of economic collusion (i.e., cartels and trusts) and other forms of anti-competitive business conduct, rent-seeking, and lobbyism, and thus the compromising of the ordoliberal competitive order and regulatory policy (Eucken, 1949).

In the context of AI, the ordoliberal (2.0) norm of transparency demands, among others, (algorithmic) explainability (i.e., 'explainable AI') and open communication (Kazim & Koshiyama, 2021), open-source data and algorithms, open government procurement, right to information, notification when an AI system makes a decision about an individual, notification when humans are interacting with AI technologies, and regular reporting (Fjeld et al., 2020).

One of the critical ethical issues, however, is that AI systems come with various problematic characteristics, including a significant degree of complexity and interconnectivity, the focus on correlation instead of causation, continuous adaptation and learning, autonomous behavior, and opacity and the 'black box problem' (Ebers, 2022). Consequently, it is often unclear how AI systems make decisions, mainly when they use machine or deep learning instead of decision trees or symbolic reasoning approaches. That is, humans are often unable to explain AI decision-making due to the opaqueness of algorithms (Pasquale, 2015), and this lack of transparency and explainability undermines society's trust in AI technologies.

Transparency – i.e., opening black box algorithms –, in turn, requires AI developers to be able to explain to affected stakeholders in everyday language how and why the model performed the way it did in a specific context (i.e., content clarification and intelligibility or explicability, which are especially important in the context of large language models and other forms of generative AI [Bender et al., 2021; EPRS, 2023; Floridi, 2023]). It is also vital to justify the ethical permissibility of AI systems, ensure their discriminatory non-harm, and generate public trust about both outcomes and processes behind the design and use. Transparency thus refers to process *and* outcome transparency and professional *and* institutional transparency – all of which are crucial for ordoliberals. Besides, it relates to professional (ordoliberal[8]) values such as integrity, honesty, neutrality, objectivity, and impartiality and the fiduciary duties of organizations towards the public and the common good (Leslie, 2019).[9]

Attard-Frost et al. adds that transparency not only refers to the scope of decision-making explanation (i.e., explaining the rationale for AI-related business decisions, especially towards impacted stakeholders) but also to transparent business practices and corporate cultures, documentation, disclosure, and selective transparency (i.e., sharing research findings and best practices and public disclosure of information, e.g., regarding the mitigation of algorithmic discrimination and biases) (Attard-Frost et al., 2022).

---

[8] Ordoliberals consider science and academics (i.e., 'clercs') as a potential ordering power in society (Eucken, 1952/2004; Röpke, 1944/1949). In the context of AI, academic researchers play an essential role in spreading digital literacy and user awareness; besides, they bear special responsibilities regarding addressing and mitigating the societal impacts of AI technologies and promoting the common good.

[9] According to Fjeld et al. (2020), professional responsibility includes accuracy, responsible design, consideration of long-term effects, multi-stakeholder collaboration, and scientific integrity (i.e., following corporate or professional codes of ethics and standards, such as a Hippocratic oath for data scientists and computer professionals).

Critical in the context of transparent or explainable AI is the concept of 'ethics by design' (Brey & Dainow, 2020; Delecraz et al., 2022; Kazim & Koshiyama, 2021; Spiekermann & Winkler, 2020; World Economic Forum, 2020). Ethics by design – or value-sensitive design – is similar to privacy by design. It deals, among others, with the so-called value-selecting (i.e., which values should be programmed into AI) and value-loading problems (how to program values into 'value-learning AI') (Wörsdörfer, 2018) and aims to increase accountability, transparency, and responsibility of AI tools, e.g., via ensuring traceability at all stages and creating 'ethical black boxes.' These devices are similar to the black boxes installed in planes and record everything an AI system does.

*Accountability and Responsibility*

One of Eucken's (1952/2004) core Constituent Principles is the liability principle. It demands that property owners (i.e., businesses and managers) be held accountable for economic decision-making, companies' business practices, and their societal impacts. In addition, managers and enterprises possess moral obligations and social responsibilities, especially toward (locally) affected stakeholder groups.[10]

In the context of AI, accountability refers to verifiability, replicability, evaluation and assessment requirements, the creation of an oversight body, the ability to appeal, remedy for automated decisions, the principle of liability and legal responsibility, and the adoption of new regulations (Fjeld et al., 2020; European Commission, 2022b; Kazim & Koshiyama, 2021; Laux, 2023). Furthermore, algorithmic accountability relates to the public perception of AI business practices (i.e., strengthening consumer confidence to get public buy-in), internal and external monitoring of AI business practices (i.e., human rights due diligence, social impact assessments and audits, internal review boards, ethics hotlines and other ethics bodies, worker involvement in the testing and review of AI system development practices, independent review processes, and authorities holding AI operators accountable, e.g., via an adequate [ordoliberal] regulatory framework) (Attard-Frost et al., 2022).

---

[10] I.e., private property is linked to and has to serve the public good (Eucken, 1952/2004); see for the social commitments of property owners: Bonhoeffer Kreis, 1979; Eucken, 1953 (here, Eucken defines private property as an essential instrument in both economic *and* social terms and stresses the socio-economic and political responsibilities of entrepreneurs).

'Accountability by design' is essential because it facilitates end-to-end answerability and auditability. It requires responsible humans-in-the-loop across the entire design and implementation chain and activity monitoring protocols that enable end-to-end oversight and review, e.g., via end-to-end recording and activity monitoring protocols (i.e., safe engineering design requires ensuring that AI systems remain under human control). Crucial are anticipatory or ex-ante accountability *and* remedial or ex-post accountability and explicability (Leslie, 2019).

According to Novelli et al. (2023), 'accountability as answerability' comes with the expectation that AI system designers, developers, and deployers comply with the respective standards and legislation to ensure the proper functioning of said technologies. Accountability rules thus introduce oversight and transparency mechanisms throughout the lifecycle of an AI system and, therefore, play a central role in fostering public trust in AI technologies – and bridging the gap between AI ethics principles and business practices (Laux, 2023).[11]

Critical in this regard is that AI systems are auditable and that certification and licensing occur: According to Mökander and Floridi (2021), ethics-based auditing would task third-party auditors with assessing whether the safety, security, privacy, and fairness-related claims made by AI developers and companies are accurate. It thus helps identify, visualize, and communicate which normative values are embedded in an AI system. Ethics-based auditing comes in three forms: functionality audits, which focus on the rationale behind the decision; code audits, which entail reviewing the source code; and impact audits, which investigate the effects of an algorithm's outputs. By promoting procedural regularity and strengthening institutional trust, ethics-based auditing can help with providing decision-making support for companies and society by visualizing and monitoring outcomes, informing individuals and society why a decision was reached and how to contest it, allowing for a sector-specific approach to AI governance, relieving human suffering by anticipating and mitigating harms, allocating accountability and responsibility by tapping into existing governance structures, and balancing actual and potential conflicts of interest, e.g., by containing access to sensitive information to an authorized third party. To ensure success, ethics-based auditing must be continuous, holistic, dialectic, strategic, and design-driven and consider the conceptual, technical, socio-economic, organizational, and institutional constraints (Mökander & Floridi, 2021).

---

[11] Novelli et al. (2023) distinguish between various accountability conditions (authority recognition, interrogation, and limitation of power), features (context, range, agent, forum, standard, process, and implications), and goals (compliance, report, oversight, and enforcement), as well as between proactive ('accountability as virtue' with the intent to prevent failures) and reactive accountability ('accountability as a mechanism' to redress failures).

According to Morley et al. (2021), the goal must be to bring ethical guidance down to the design level by providing tools and methods that translate AI ethics into technical specifications, i.e., creating a bridge between abstract principles and technical implementation, similar to the Digital Catapult AI Ethics Framework. Critical are regular ethical evaluations as an integral part of a system's operation (i.e., validation, verification, and evaluation) and avoiding one-off tick-box exercises completed only at the beginning of the design process. External algorithmic audits – e.g., conducted by Aequitas and Turingbox – are critical to operationalizing AI ethics. Yet, we must be aware of the limits of such audits, e.g., they only provide us with an ex-post evaluation, i.e., after the system has been marketed. Furthermore, the agile nature of machine/deep learning poses additional challenges for AI audits. Internal audits with full access to data (i.e., code review) and a focus on ethical foresight could help to address those issues. They can check for reliability and robustness and include risk anticipation and mitigation. The main problem with internal audits is that they might run into conflicts of interest and lack objectivity. What is needed is thus a multi-agent system with distributed responsibilities across different agents, i.e., 'ethics as a service.' Ethics as a service relies on distributed responsibility and is based on platform as a service – similar to cloud computing (i.e., software, infrastructure, and platform as a service). Its key components include independent multi-disciplinary ethics advisory boards, collaboratively developed ethical codes, and the involvement of AI practitioners (Morley et al., 2021).

*Democracy and the Rule of Law*

According to Eucken and Böhm (1966/1980), a genuinely ordoliberal society requires eliminating all social privileges, especially those related to class and status. It also requires measures to overcome all remaining 'plutocratic elements' of a 'privilege-based feudal society.' The ideal would be a (Kantian) society where all members possess de facto equal rights and opportunities and egalitarian status as 'private law subjects.' In such an ideal ordoliberal society, all members could coordinate their plans – and satisfy their preferences and needs – with the help of competitive market processes – instead of the usual dependencies, oppression, and subordination in feudal societies. By transforming a privilege-based into a private law society (Wörsdörfer, forthcoming), a coexistence of free, equal (in terms of legal status), and autonomous individuals would emerge. For this to happen, private autonomy must be codified and enshrined in the constitution and embedded in a democratic, rule-of-law society. Here, citizens would have their fundamental human rights protected by a constitutional state focusing solely on regulatory rather than process policy. In short, Eucken and Böhm's private law society rests on the guarantee of private property, the protection of freedom of contract (as long as it is compatible with a competitive market economy), a 'market order society' (which allows for the coordination of individual plans instead of subordination), and the interdependency of the legal and economic order (i.e., private law and market economy) (Vanberg, 2005, 2013). Crucial elements of the ordoliberal private law society are freedom, competition, and regulatory policy. In the context of AI, Eucken and Böhm's private law society would require embedding AI systems – and the digital economy in general – in democratic and rule-of-law societies, with adequate parliamentary and judicial oversight, similar to Häußermann and Lütge's (2022) concept of 'deliberative order ethics' (this is especially the case for generative AI, including chatbots and the underlying large language models [Bender et al., 2021; EPRS, 2023; Floridi, 2023]). The concept is based on a contractualist theory of business ethics that builds on ordoliberalism and rests on participation (i.e., participatory democracy) and deliberation (i.e., public debates), i.e., inclusive, equal, and diverse stakeholder dialogue and engagement processes – similar to Habermas' discourse ethics –, and the so-called community-in-the-loop approach. According to the authors, deliberative order ethics can address the shortcomings of current AI ethics initiatives, i.e., the neglect of the importance of business practices, the bias towards technological solutionism, the focus on individuals instead of society, the lack of practical implementation, accountability, and clear impact, and the unclear relationship between AI ethics and regulation.

*Environmental and Social Sustainability*

Eucken's Regulating Principles (1952/2004) demand, inter alia, the correction of adverse external effects and the internalization of social costs, such as environmental pollution and other forms of ecological damage. He also stresses several times not only the environmental obligations of businesses and managers but also their social responsibilities and common good orientation.

In the context of AI, environmental sustainability relates to the ecological impacts and carbon footprint of AI technologies, such as the significant energy consumption and corresponding greenhouse-gas emissions of data centers or the problem of electronic waste. Noteworthy is that the E.U.'s AIA (see Section 4), the gold standard in AI legislation, does not adequately address systemic sustainability risks created by AI systems, and there are no direct references to climate change, ecological sustainability, environmental protection, and green or sustainable AI – except in the context of voluntary codes of conduct (Ebers et al., 2021; EPRS, 2022a). As Floridi (2021) and others have pointed out, the E.U. primarily follows an anthropocentric approach, prioritizing human needs at the expense of environmental concerns. This, however, could be overcome with a Euckenian framework focusing on correcting adverse external effects and promoting ecological sustainability (e.g., with the help of carbon taxes or emissions trading schemes).

On the other hand, social sustainability requires AI developers to proceed with continuous sensitivity to real-world impacts; technical sustainability, particularly, depends on product safety (i.e., accuracy, reliability, security, and robustness). Crucial in this regard is human rights due diligence and stakeholder impact assessment, i.e., assessing the possible impacts on individuals, society, and interpersonal relationships (Leslie, 2019). Additionally, social sustainability refers to sustainable development (i.e., education, training, and development of the AI workforce are crucial for companies to innovate successfully) and the management and distribution of societal benefits and harms (Attard-Frost et al., 2022).

An overview of the discussed ordoliberal-inspired AI ethics principles can be found in Figure 2.

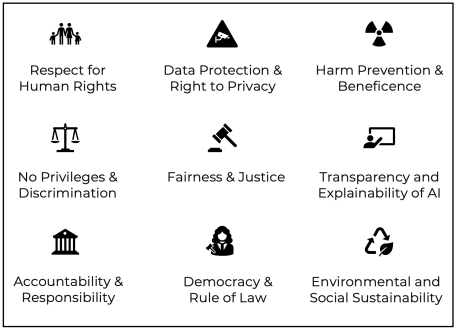

*Figure 2: Ordoliberal AI Ethics Principles*

## 4. An Ordoliberal Framework Policy for AI

From an ordoliberal 2.0 perspective, implementation of the above principles rests on two pillars – regulatory policy (i.e., adequate AI legislation) and competition policy (i.e., big tech and antitrust). The following paragraphs analyze both pillars and show how they could help realize the aforementioned ordoliberal-inspired AI ethics principles.

*Regulatory Policy (AI Legislation)*

The E.U.'s Artificial Intelligence Act (AIA) is a crucial step in the right direction to realize the first pillar of an ordoliberal framework for AI ethics (European Commission, 2021a, 2021b, 2021c, 2022a). The AIA's primary goal is to create a legal framework for secure, trustworthy, and ethical AI. I.e., it aims to ensure that AI technologies are human-centric, safe to use, compliant with the law, and respectful of fundamental rights, especially those enshrined in the Charter of Fundamental Rights (see *Principle 1* above). The AIA results from extensive stakeholder consultation, incorporates input from the High-Level Expert Group on AI (2019), and builds on the E.U. White Paper (European Commission, 2020) and other AI ethics initiatives. It is part of the E.U.'s digital single market strategy, complements the General Data Protection Regulation (see *Principle 2*), and is consistent with the Digital Services Act, DMA, and other regulatory initiatives such as the Data (Governance) Act, the (revised) AI Liability Directive(s), and sectoral product safety frameworks (Dheu et al., 2022; Hacker et al., 2023; Wörsdörfer, 2023a, 2023b, 2023c).

The AIA sorts AI systems into different risk categories – unacceptable risk, high risk, and limited or minimal risk categories (the higher the risk, the more regulations apply to those technologies). It denies market access whenever the risks are deemed too high for risk-mitigating interventions (i.e., prohibited AI systems). For high-risk AI systems, market access is granted if they comply with the AIA, i.e., the AIA imposes ex-ante technical requirements and an ex-post market monitoring procedure on those systems; minimal-risk systems need to fulfill general safety requirements, such as the ones included in the General Product Safety Directive. In short, the AIA defines prohibited AI practices, such as social scoring, and high-risk AI systems, that is, systems that pose significant risks to the health and safety of individuals or negatively affect the fundamental rights of persons (note that based on the Parliament's recommendations, chatbots and deepfakes would be considered high-risk [European Parliament, 2023a, 2023b, 2023c]). It also defines areas where (real-time) facial recognition (Almeida et al., 2021) is allowed, restricted, and prohibited and imposes transparency and other obligations on high-risk AI technologies (see *Principle 6*). With the AIA, the E.U. recognizes the adverse effects of AI technologies on fundamental rights and safety and that voluntary self-regulation offers inadequate protection. The Commission thus changed course – from soft to hard law – and reversed its previous strategy, which was based on the recommendations of the High-Level Expert Group and the White Paper. The AIA also attempts to foster innovation – e.g., with the so-called regulatory sandboxes (EPRS, 2022b) – while at the same time protecting humans, the rule of law, and democracy (see *Principle 8*). Lastly, it tries to create a level playing field of protection across E.U. member states and prioritizes respect for fundamental rights (see *Principle 1*), including health and safety (Smuha et al., 2021). At the time of writing, the revised AIA proposal has been approved by the Parliament, and negotiations with the Council are expected to start soon. Both institutions – and the Commission – must agree on a common text before the AIA can be adopted, which is expected to be in 2024. Among the AIA's strengths are its legally binding, i.e., hard-law, character (Ebers, 2020, 2022), which marks a welcoming departure from existing soft-law AI ethics initiatives (Attard-Frost et al., 2022; EPRS, 2022a; Fjeld et al., 2020; Leslie, 2019; Leslie et al., 2021; Mittelstadt, 2019; Rubenstein, 2021). Other positive aspects – as seen from an ordoliberal 2.0 point of view – include the AIA's extra-territoriality and possible extension of the 'Brussels Effect' (Bradford, 2020; Floridi, 2021; Greenleaf, 2021; Petit, 2020), the ability to address data quality and discrimination risks (Hacker, 2021), and institutional innovations such as the European Artificial Intelligence Board (EAIB) and publicly accessible logs and database for AI systems, as an essential step in opening black-box algorithms (AlgorithmWatch, 2021). Yet from an ordoliberal 2.0 perspective, the AIA falls short of realizing its full potential: Ordoliberals are primarily concerned with the AIA's proposed governance structure; they specifically criticize its lack of effective enforcement, oversight and control mechanisms, procedural rights, worker protection, institutional clarity, sufficient funding and staffing, and consideration of sustainability issues (AlgorithmWatch, 2021; Almada & Petit, 2023; Biber, 2021; Castets-Renard & Besse, 2022; Ebers et al., 2021; EPRS, 2022a, 2022c; Floridi, 2021; Gstrein, 2022; Kazim et al., 2022; Mahler, 2022;

Mazzini & Scalzo, 2022; Mökander et al., 2022; Smuha et al., 2021; Stuurman & Lachaud, 2022; Veale & Zuiderveen Borgesius, 2021; Wachter et al., 2021; Wörsdörfer, 2023b, 2023c).[12]

To address these issues and bring the AIA closer to an alignment with the ordoliberal (2.0) ideal, several reform measures (Wörsdörfer, 2023b, 2023c) need to be taken, such as introducing or strengthening …

- *Conformity assessment procedures*: The AIA needs to move beyond the currently flawed system of provider self-assessment and certification towards mandatory third-party audits for all high-risk AI systems; i.e., the existing governance regime, which involves a significant degree of discretion for self-assessment and certification for AI providers and technical standardization bodies, needs to be replaced with legally mandated external oversight by an independent regulatory agency with appropriate investigatory and enforcement powers (AlgorithmWatch, 2021; Ebers et al., 2021; EPRS, 2022a; Floridi et al., 2022; Gstrein, 2022; Kop, 2021; Laux et al., 2023b).

- *Democratic accountability and judicial oversight* (see *Principle 7*): What is needed is a meaningful engagement of all affected groups, including consumers and social partners (e.g., workers exposed to AI systems and unions[13]), and a public representation in the context of standardizing and certifying AI technologies. The overall (ordoliberal) goal is to ensure that those with less bargaining power are included and their voices are heard (Ebers et al., 2021; EPRS, 2022a, 2022c; Gstrein, 2022).[14]

---

[12] Other points of criticism include the AIA's tendency to prioritize economic, business, and innovation over moral concerns (i.e., the de-prioritization of human rights), the lack of a clear definition of AI systems (i.e., a lack of scope), the flawed risk-based framework (i.e., an incomplete list of prohibited AI systems and under-regulation of non-high-risk AI systems), and the failure to adequately address the challenges posed by generative AI, such as chatbots and deepfakes (EPRS, 2023; Floridi, 2023; Gebru et al., 2023).

[13] Note that Eucken and other ordoliberals saw unions as an essential counterweight to the power of employers (Eucken, 1952/2004, 1999).

[14] Experts recommend making the risk classification and standardization process more inclusive and transparent (Ebers, 2022). This would require substantive information rights for affected individuals, adding public participation rights for citizens, and ensuring that not only corporate and expert groups are involved in the classification and standardization process by actively involving organizations that represent public interests (Smuha et al., 2021). It might also be worth exploring whether the EAIB's responsibilities could be expanded. Currently, the board serves in an advisory role, and critics claim that its tasks should be amended to include investigatory and regulatory powers. Furthermore, it could be transformed into a stakeholder forum to overcome the previously mentioned issues of lack of consultation, participation, and stakeholder dialogue (Council of the European Union, 2022a, 2022b). Besides reforming the AIA, democratic accountability and judicial oversight require an ordoliberal-inspired competition policy and a strengthening of the DMA (see below).

- *Redress and complaint mechanisms*: Besides consultation and participation rights, ordoliberals also request the inclusion of explicit information rights, easily accessible, affordable, and effective legal remedies, and individual and collective complaint and redress mechanisms. That is, bearers of fundamental rights must have means to defend themselves if they feel they have been adversely impacted by AI systems or treated unlawfully. I.e., AI subjects must be able to legally challenge the outcomes of such systems (AlgorithmWatch, 2021; EPRS, 2022a, 2022c; Smuha et al., 2021).
- *Worker protection*: Ordoliberals demand better involvement and protection of workers and their representatives in using AI technologies. This could be achieved by classifying more AI at-work systems as high-risk or prohibiting them. Workers should also be able to participate in management decisions regarding using AI tools in the workplace. Their voices and concerns should be heard, especially when technologies that might negatively impact their work experience are introduced. Workers should, moreover, have the right to object to the use of specific AI tools in the workplace and file complaints (see *Principle 1*) (AlgorithmWatch, 2021; Cefaliello & Kullmann, 2022; EPRS, 2022c; Smuha et al., 2021).
- *Governance structure*: Effective enforcement of the AIA also hinges on strong institutions and 'ordering powers' (Eucken, 1952/2004; Röpke, 1942, 1944/1949, 1950; Rüstow, 1955, 1957, 2001). The EAIB has the potential to be such a power and to strengthen AIA oversight and supervision. This, however, requires that it has the corresponding capacity, expertise (in both technology and fundamental rights), resources, and political independence. To ensure adequate transparency – as demanded by ordoliberalism (see *Principle 6*) – the E.U.'s AI database should include not only high-risk systems but *all* forms of AI technologies. Moreover, it should list all systems used by private *and* public entities. The material provided to the public should include information regarding algorithmic risk and human rights impact assessment. This data should be available to those affected by AI systems in an easily understandable and accessible format (AlgorithmWatch, 2021; Ebers et al., 2021; Smuha et al., 2021).
- *Funding and staffing of market surveillance authorities*: Besides the EAIB and AI database, national authorities must be strengthened – both financially and expertise-wise. It is worth noting that the 25 full-time equivalent positions foreseen by the AIA for national supervisory authorities (see AIA Impact Assessment, Annex 3 [European Commission, 2021d]) are insufficient and that additional financial and human resources must be invested in regulatory agencies to effectively implement the proposed AI regulation (AlgorithmWatch, 2021; EPRS, 2022c; Smuha et al., 2021).

- *Sustainability considerations* (see *Principle 9*): To better address the adverse external effects and environmental concerns of AI systems, ordoliberals also demand the inclusion of sustainability requirements for AI providers, e.g., obliging them to reduce the energy consumption and e-waste of AI technologies, thereby moving towards green AI. Ideally, those requirements should be mandatory and go beyond the existing (voluntary) codes of conduct (AlgorithmWatch, 2021; Ebers et al., 2021; EPRS, 2022a).[15]

*Competition Policy (Big Tech and Antitrust)*

AI legislation is essential in realizing the above ordoliberal AI ethics principles. Yet, from an ordoliberal 2.0 point of view, it needs to be accompanied by an adequate competition policy to address the power asymmetries in the digital economy, particularly between big tech and consumers and between big tech and small and medium-sized companies.

As Wörsdörfer (2022b) has shown, the current antitrust regimes of the E.U. and especially the U.S. are flawed. They are, among others, unable to (fully) realize a competitive economy, open up markets – and ensure low market entry and exit barriers –, correct market power, limit lobbyism and rent-seeking, adequately review and block mergers and acquisitions (M&As), and implement behavioral and structural remedies, as envisioned by ordoliberalism 2.0. The E.U. has shown promising potential, especially with its recent antitrust probes (e.g., against app store, mobile payment, and IoT service providers) and policy proposals (e.g., the DMA). Yet, to fully realize the above ordoliberal 2.0 criteria, the E.U.'s antitrust regime needs to be further strengthened, including hardening the DMA (Wörsdörfer, 2023a). This could be achieved with the help of the following ordoliberal-inspired reform proposals (Wörsdörfer, 2022b):

---

[15] Note that an ordoliberal-inspired social and environmental market economy would mandate the internalization of adverse external effects (e.g., via carbon pricing or emissions trading schemes) and would thus help reduce energy usage – including the ones of AI systems – and promote green environments.

- *Updating antitrust laws and making them fit for the digital economy*: I.e., passing new rules and regulations specifically designed for digital platforms; it also implies updating market concepts and definitions, developing new analytical tools, and considering new forms of market power, e.g., data power, digital nudging power, bottleneck power, strategic market status power, and gatekeeping or rule-setting power.
- *Shifting the burden of proof from competition agencies to the merging parties*: I.e., new laws and regulations would shift the burden of proof to the involved parties, which would need to prove the benefits of M&As or at least that the M&A does not harm, i.e., reduce competition or that M&A-specific efficiencies for the public offset the possible adverse effects on competition (see *Principle 3*); such a shift would make it easier to initiate and prove antitrust cases.
- *Establishing an anti-merger presumption*: I.e., all M&As that could lead to more concentrated markets are considered illegal – unless the merging parties can prove otherwise; such a presumptive prohibition is fundamental in the context of dominant digital platforms and powerful gatekeepers; that is, future M&As involving big tech could be presumed anti-competitive unless the involved parties can prove their societal benefits (see *Principle 3*).
- *Revising existing merger guidelines and introducing ex-post merger control*: What is crucial yet missing in both the U.S. and E.U. is to establish a look-back mechanism or an ex-post merger control; that is, antitrust agencies should be able to conduct regular retrospective reviews of previous M&As and reexamine their medium- and long-term impacts on the economy. The empirical assessment would specifically focus on whether past M&As have been in the public interest or harmed competition, e.g., via exclusionary and discriminatory business conduct. Suppose competition authorities determine that anti-competitive harm has occurred. In that case, they should be able to take measures that would help remedy any corresponding damage, including a possible demerger or merger unwinding (see *Principle 3*).
- *Making more frequent use of behavioral and structural remedies*: Behavioral remedies include ensuring data portability and interoperability, prohibiting predatory pricing, banning tying and bundling, and preventing self-preferencing. As a measure of last resort or ultima ratio – not only behavioral but also structural remedies could be warranted; those measures, already recommended by ordoliberals in the 1940s (Eucken, 1946/199b), could, for instance, include the separation of business units, the unbundling of companies, the horizontal division of services, and (other) changes in the financial organization of a company.

- *Ensuring platform neutrality similar to net neutrality*: Such a rule would require gatekeepers to operate their platforms neutrally, including granting equal access and fair and equitable terms and conditions to all companies and services. In short, the non-discrimination policy would prohibit big tech companies from privileging their services, and the ordoliberal neutrality principle would ensure free, open, and fair platform access (see *Principles 4 and 5*).

- *Better funding and stuffing of antitrust agencies:* To better enforce the suggested proposals, monitor business conduct, and sanction non-compliance, independent antitrust agencies must be set up (e.g., similar to the U.K.'s Digital Markets Unit) and better funded and staffed (e.g., by hiring data scientists, AI specialists, and independent researchers).

- *Increasing monetary penalties for anti-competitive business practices*: Besides enforcement and monitoring, good governance also hinges on adequate sanctioning mechanisms. Ordoliberals, therefore, recommend significantly increasing the penalties and fines against cartels and for companies using exclusionary business practices – in addition to behavioral and structural remedies (together, they could function as an ex-ante deterrent).

- *Better protection of whistle-blowers*: For an antitrust lawsuit to succeed, agencies must receive first-hand or insider information. This, however, requires better protection of whistleblowers and incentivizing them to share their information and knowledge, e.g., about cartels and anti-competitive business conduct, with the respective antitrust authorities. This could be achieved with the help of a reward or bounty system for individuals exposing anti-competitive or exclusionary practices or cartels.

- *International cooperation*: National antitrust efforts must be complemented by an international ordoliberal agenda for promoting competition in the digital era. This could, for instance, entail cross-border cooperation between various antitrust agencies, national governments, and international organizations (e.g., OECD, G7, and G20). The primary purpose of this form of collaboration is to share best practices, guidelines, tools, and information and develop a common (i.e., unified) approach toward digital platforms. What needs to be avoided is a fragmentary regulatory landscape, non-coherent measures, and a 'race to the bottom'; instead, governments and international organizations should strive towards increased harmonization and a 'race to the top' (this last reform proposal is also crucial in the context of AI legislation, thereby linking the regulatory and competition policy narrative).

Figure 3 shows how the two pillars – regulatory policy (AI legislation) and competition policy (big tech and antitrust) – relate to ordoliberalism 2.0 and the ordoliberal-inspired AI ethics principles.

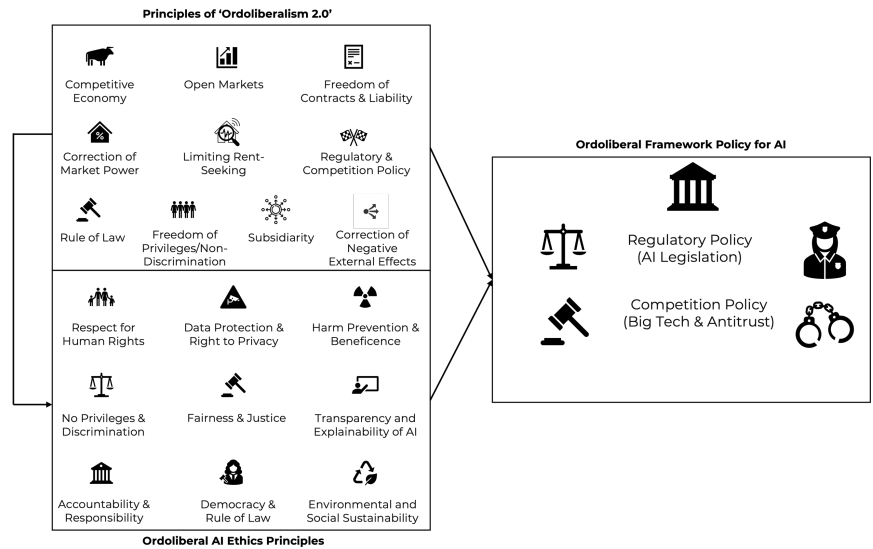

*Figure 3: Ordoliberal Framework Policy for AI*

## 5. Concluding Remarks

This paper has summarized ordoliberalism and its key characteristics (Section 2). It also identified several ordoliberal-inspired AI ethics principles that could serve as the foundation for a 'digital bill of rights,' namely respect for human rights, data protection and the right to privacy, harm prevention and beneficence, non-discrimination and freedom of privileges, fairness and justice, transparency and explainability of AI systems, accountability and responsibility, democracy and the rule of law, and environmental and social sustainability (Section 3). Implementing those AI ethics principles requires an accompanying ordoliberal-inspired regulatory (i.e., effective AI legislation and reformed AIA) and competition policy (i.e., effective counter-measures against the anti-competitive business conduct of big tech). As Section 4 has shown, both policy areas have yet to receive the attention they deserve; this is especially the case for the intersection or combination of AI ethics, regulation, and antitrust. Therefore, fundamental reform is needed to align the above-described ordoliberal ideals with current policy proposals and measures.

Our research has, so far, focused on the macro level (i.e., nation-states and governments, including the E.U.). In the next step of our research project, we try to broaden our approach and incorporate the micro (i.e., data scientists and AI researchers) and meso (i.e., corporations and organizations) levels. We specifically aim to provide concrete guidance for AI developers, operators, and policy advisors – going above and beyond the general overview of reform measures presented in this paper – in the hope that some of the suggested policy proposals will find their way into upcoming AI ethics and antitrust legislations as well as professional and corporate codes of conduct around the globe.

Funding: *The author(s) declare(s) that no funds, grants, or other support were received during the preparation of this manuscript.*

*Conflicts of interest: The author(s) has/have declared no conflict of interest/The author(s) has/have no relevant financial or non-financial interests to disclose.*